\documentclass[12pt]{article}
\begin{document}
\newcommand{\Robs}{{\cal R}}
\newcommand{\spur}{{\rm Sp}\,}
\newcommand{\M}{{\cal M}}
\newcommand{\bra}{\langle}
\newcommand{\ket}{\rangle}
\newcommand{\eq}[2]{\begin{equation}\label{#1} #2 \end{equation}}
\newcommand{\HH}{{\cal H}}
\newcommand{\W}{{\omega}}
\newcommand{\then}{\Longrightarrow}
\newcommand{\spr}[2]{\sigma(#1,\, #2)}
\newcommand{\tbf}[1]{{\bf #1}}
\newcommand{\tit}[1]{{\it #1}}
\newcommand{\cl}[1]{{\cal #1}}
\newcommand{\half}{{\scriptstyle \frac12}}
\newcommand{\OPS}{\HH^{(1)}}
\newcommand{\F}{{\cal F}}
\newcommand{\KS}{{\Sigma}}
\newcommand{\vect}[1]{\left(\begin{array}{c}#1\end{array}\right)}
\newcommand{\mat}[1]{\left(\begin{array}{cc}#1\end{array}\right)}
\newcommand{\Rcite}[1]{ref.\cite{#1}}
\newcommand{\Alg}{{\cal U}}
\newcommand{\RWA}{{\cal U}_R}
\newcommand{\LWA}{{\cal U}_L}
\newcommand{\DWA}{\tilde{\cal U}}
\newcommand{\intRR}{\int\limits_{-\infty}^{\infty}}
\newcommand{\intR}{\int\limits_{0}^{\infty}}
\newcommand{\KMS}{{\rm KMS}}
\newcommand{\WF}{{\cal W}}
\newcommand{\LRD}[1]{\frac{{{\displaystyle\leftrightarrow}
\atop {\displaystyle\partial}}}{\partial #1}}
\newcommand{\lrd}[1]{\stackrel{\displaystyle \leftrightarrow}
{\displaystyle\partial_{#1}}}
\newcommand{\diff}[1]{\partial /{\partial #1}}
\newcommand{\Diff}[1]{\frac{\partial}{\partial #1}}
\newcommand{\hc}[1]{#1^{\dagger}}

\newcount\secnum \secnum=0
\def\newsec{\advance\secnum by 1
\vskip 0.5cm
{\tbf{\the\secnum.\quad}}\nobreak}


\title{{\bf Quantum field aspect \\
of Unruh problem}}
\author{
{\it A.M. Fedotov$^{*\,i)}$,}
{\it V.D. Mur$^*$,}
{\it N.B. Narozhny$^{*\,ii)}$,} \\
{\it V.A. Belinski\u\i$^{+\,iii)}$}
{\it and B.M. Karnakov$^*$}\\[0.3cm]
{${}^*$Moscow State Engineering Physics Institute,}\\
{115409 Moscow, Russia}\\[0.3cm]
{${}^+$INFN and ICRA, Rome University "La Sapienza",}\\
{00185 Rome, Italy}\\[2cm]}
\date{}
\maketitle
\vspace{-2cm}
\centerline{\bf Abstract}

\vspace{1cm}
It is shown using both conventional and algebraic approach
to quantum field theory that it is impossible to perform
quantization on Unruh modes in Minkowski spacetime. Such
quantization implies setting boundary condition for the
quantum field operator which changes topological properties
and symmetry group of spacetime and leads to field theory in
two disconnected left and right Rindler spacetimes. It means
that "Unruh effect" does not exist.

\vspace{0.5cm}

\parindent=0cm
PACS: 03.70.+{\bf k},04.70.Dy \\
Keywords: Quantum field theory; Fulling particles;
Unruh modes; Unruh effect

\vfill \footnoterule \leftline{$^{i)}$E-mail: fedotov@cea.ru}
\leftline{$^{ii)}$E-mail: narozhny@pc1k32.mephi.msk.ru}
\leftline{$^{iii)}$E-mail: volodia@vxmrg9.icra.it}
\thispagestyle{empty}
\newpage
\setcounter{page}{1}

\parindent 0.7cm
\newsec
The Unruh problem \cite{Unruh} closely associated with the Hawking
effect \cite{Hawk} is known for more than twenty years. It is
asserted that from the point of view of a uniformly accelerated
observer in Minkowski spacetime (MS) the usual vaccum state $|0_{\M}\ket$
looks like a mixed state described by the thermal density matrix with
Davies -- Unruh \cite{Unruh,Dav} effective temperature $T=a/2\pi$
\footnote{We use units $\hbar=1$, $c=1$.} where $a$ stands for the
proper acceleration of the observer.
More precisely (see e.g. Refs.[4-9] and citation therein) one has the
relation
\eq{main1}{\bra 0_{\M}|\Robs |0_{\M}\ket=\spur(\rho_R\Robs)
,\quad \rho_R=Z^{-1}\exp(-2\pi H_R).}
Here $\Robs$ is an arbitrary observable which depends on "values of the
field" $\phi(x)$ only for $x$ from the right Rindler wedge $R$ (see Fig.1)
which contains the world line of the observer, $\rho_R$ is the density
matrix and $H_R$ is the secondly quantized Hamiltonian with respect to
timelike variable $\eta$ in $R$.

An evristic explanation of Unruh effect is the following. Spacetime in
the Rindler reference frame (with respect to which the accelerated observer
is at rest) possesses event horizons. Therefore the Rindler observer looses
a part of information accessible for inertial observer in MS. Hence he
perceives the Minkowski vaccum state as a mixed state.

Mathematically correct consideration of Unruh problem is possible in
the frame of algebraic approach to quantum field theory (see reviews
\cite{Emch,Haag} on algebraic approach and Ref.\cite{Kay} for it's
application to Unruh problem) which allows one to consider pure and
mixed states on the unified grounds. In this approach a notion of
Kubo -- Martin -- Schwinger (KMS) state \cite{HHW} is used as a
definition of thermal equilibrium state.

In the current paper we will show that eq.(\ref{main1}) as well as it's
analog in algebraic approach cannot serve as a proof of Unruh effect.
The reason is the existance of boundary condition \cite{last} for the
field operator in Rindler spacetime (RS). We will also point out a
generalization of this boundary condition in algebraic treatment.
But first we will discuss an equivalent interpretation of the Unruh
effect in terms of Fulling -- Unruh "particles". The latter arises
when the so -- called Unruh modes \cite{Unruh,UW} are used for
quantization of the field. We will show that the Unruh modes can be
used as a basis for quantization only in double RS rather than in MS.


\parindent 0.7cm
\newsec
Since the Rindler observer world line coincides with one of the orbits
of Lorentz rotation we consider quantization of neutral scalar field
\footnote{We restrict ourselves to the case of two dimentional
spacetime. This assumpltion is choosen only to simplify
notation and does not affect the results.}
in the basis of boost generator eigenfunctions
\eq{boost}{
\Psi_{\kappa}(t,z)=2^{-3/2}\pi^{-1}\intRR d\theta\,
\exp\{-im(t\cosh\theta-z\sinh\theta)-i\kappa\theta\}.}
These modes are positive frequency with respect to global Minkowski
time solutions of the Klein -- Gordon (KG) equation. They are orthonormal
relative to the KG inner product
$$\bra f,g\ket_{\M}=i\intRR dz\,f^*(x)\LRD{t} g(x),
\quad x=(t,z),$$
form a complete set and hence may be used for the field quantization
\footnote{Quantization of scalar field performed in Ref.\cite{Boul}
by analitical continuation of Green functions is equivalent to the
one defined by Eqs.(\ref{boost}), (\ref{BMquant}).},
\eq{BMquant}{
\begin{array}{c}\displaystyle
\phi(x)=\intRR d\kappa\,\{\Psi_{\kappa}(x)b_{\kappa}+
\Psi_{\kappa}^*(x)\hc{b_{\kappa}} \} \\
\displaystyle
[b_{\kappa},\hc{b_{\kappa'}}]=\delta(\kappa-\kappa'),\quad
b_{\kappa}|0_{\M}\ket=0,\quad -\infty<\kappa<\infty.
\end{array}}
For $b_{\kappa}$ we have
\eq{b1}{b_{\kappa}=\bra\Psi_{\kappa},\phi\ket_{\M}.}

\begin{figure}
\parindent=-1cm
%
\setlength{\unitlength}{0.240900pt}
\ifx\plotpoint\undefined\newsavebox{\plotpoint}\fi
\sbox{\plotpoint}{\rule[-0.175pt]{0.350pt}{0.350pt}}%
\begin{picture}(1500,1530)(0,0)
\sbox{\plotpoint}{\rule[-0.175pt]{0.350pt}{0.350pt}}%
\put(100,788){\vector(1,0){1500}} \put(850,100){\vector(0,1){1500}}
\put(800,1550){$t$} \put(1550,738){$z$}
\put(850,788){\line(2,1){600}} \put(1400,1000){$\eta={\rm const}$}
\put(1300,400){$\rho={\rm const}$}
\put(1300,405){\vector(-1,0){60}}
\put(850,1200){\Large F} \put(850,400){\Large P}
\put(1200,700){\Large R} \put(400,700){\Large L}
\put(1340,1440){$h_-^+$} \put(300,1440){$h_+^-$}
\put(1340,100){$h_+^+$} \put(300,100){$h_-^-$} \put(740,740){$h_0$}
\put(790,760){\vector(2,1){50}}
\sbox{\plotpoint}{\rule[-0.500pt]{1.000pt}{1.000pt}}%
\put(264,1397){\usebox{\plotpoint}}
\put(265,1394){\usebox{\plotpoint}}
\put(267,1392){\usebox{\plotpoint}}
\put(269,1390){\usebox{\plotpoint}}
\put(271,1388){\usebox{\plotpoint}}
\put(273,1386){\usebox{\plotpoint}}
\put(275,1384){\usebox{\plotpoint}}
\put(277,1381){\usebox{\plotpoint}}
\put(279,1379){\usebox{\plotpoint}}
\put(281,1377){\usebox{\plotpoint}}
\put(283,1375){\usebox{\plotpoint}}
\put(285,1373){\usebox{\plotpoint}}
\put(287,1371){\usebox{\plotpoint}}
\put(289,1368){\usebox{\plotpoint}}
\put(291,1366){\usebox{\plotpoint}}
\put(293,1364){\usebox{\plotpoint}}
\put(295,1362){\usebox{\plotpoint}}
\put(297,1360){\usebox{\plotpoint}}
\put(299,1358){\usebox{\plotpoint}}
\put(301,1355){\usebox{\plotpoint}}
\put(303,1352){\usebox{\plotpoint}}
\put(305,1350){\usebox{\plotpoint}}
\put(307,1347){\usebox{\plotpoint}}
\put(309,1345){\usebox{\plotpoint}}
\put(311,1342){\usebox{\plotpoint}}
\put(313,1340){\usebox{\plotpoint}}
\put(315,1338){\usebox{\plotpoint}}
\put(317,1336){\usebox{\plotpoint}}
\put(319,1334){\usebox{\plotpoint}}
\put(321,1332){\usebox{\plotpoint}}
\put(323,1329){\usebox{\plotpoint}}
\put(325,1327){\usebox{\plotpoint}}
\put(327,1325){\usebox{\plotpoint}}
\put(329,1323){\usebox{\plotpoint}}
\put(331,1321){\usebox{\plotpoint}}
\put(333,1319){\usebox{\plotpoint}}
\put(335,1316){\usebox{\plotpoint}}
\put(337,1314){\usebox{\plotpoint}}
\put(339,1312){\usebox{\plotpoint}}
\put(341,1310){\usebox{\plotpoint}}
\put(343,1308){\usebox{\plotpoint}}
\put(345,1306){\usebox{\plotpoint}}
\put(347,1303){\usebox{\plotpoint}}
\put(349,1301){\usebox{\plotpoint}}
\put(351,1299){\usebox{\plotpoint}}
\put(353,1296){\usebox{\plotpoint}}
\put(355,1294){\usebox{\plotpoint}}
\put(357,1292){\usebox{\plotpoint}}
\put(359,1289){\usebox{\plotpoint}}
\put(361,1287){\usebox{\plotpoint}}
\put(363,1285){\usebox{\plotpoint}}
\put(365,1283){\usebox{\plotpoint}}
\put(367,1281){\usebox{\plotpoint}}
\put(369,1279){\usebox{\plotpoint}}
\put(371,1276){\usebox{\plotpoint}}
\put(373,1274){\usebox{\plotpoint}}
\put(375,1272){\usebox{\plotpoint}}
\put(377,1269){\usebox{\plotpoint}}
\put(379,1267){\usebox{\plotpoint}}
\put(381,1265){\usebox{\plotpoint}}
\put(383,1262){\usebox{\plotpoint}}
\put(385,1260){\usebox{\plotpoint}}
\put(387,1258){\usebox{\plotpoint}}
\put(389,1255){\usebox{\plotpoint}}
\put(391,1253){\usebox{\plotpoint}}
\put(393,1251){\usebox{\plotpoint}}
\put(395,1248){\usebox{\plotpoint}}
\put(397,1246){\usebox{\plotpoint}}
\put(399,1244){\usebox{\plotpoint}}
\put(401,1242){\usebox{\plotpoint}}
\put(403,1240){\usebox{\plotpoint}}
\put(405,1238){\usebox{\plotpoint}}
\put(407,1235){\usebox{\plotpoint}}
\put(409,1233){\usebox{\plotpoint}}
\put(411,1231){\usebox{\plotpoint}}
\put(413,1228){\usebox{\plotpoint}}
\put(415,1226){\usebox{\plotpoint}}
\put(417,1224){\usebox{\plotpoint}}
\put(419,1221){\usebox{\plotpoint}}
\put(421,1219){\usebox{\plotpoint}}
\put(423,1217){\usebox{\plotpoint}}
\put(425,1215){\usebox{\plotpoint}}
\put(427,1213){\usebox{\plotpoint}}
\put(429,1211){\usebox{\plotpoint}}
\put(431,1208){\usebox{\plotpoint}}
\put(433,1206){\usebox{\plotpoint}}
\put(435,1204){\usebox{\plotpoint}}
\put(437,1201){\usebox{\plotpoint}}
\put(439,1199){\usebox{\plotpoint}}
\put(441,1197){\usebox{\plotpoint}}
\put(443,1194){\usebox{\plotpoint}}
\put(445,1191){\usebox{\plotpoint}}
\put(447,1189){\usebox{\plotpoint}}
\put(449,1186){\usebox{\plotpoint}}
\put(451,1184){\usebox{\plotpoint}}
\put(453,1181){\usebox{\plotpoint}}
\put(455,1179){\usebox{\plotpoint}}
\put(457,1177){\usebox{\plotpoint}}
\put(459,1175){\usebox{\plotpoint}}
\put(461,1173){\usebox{\plotpoint}}
\put(463,1171){\usebox{\plotpoint}}
\put(465,1168){\usebox{\plotpoint}}
\put(467,1166){\usebox{\plotpoint}}
\put(469,1164){\usebox{\plotpoint}}
\put(471,1161){\usebox{\plotpoint}}
\put(473,1159){\usebox{\plotpoint}}
\put(475,1157){\usebox{\plotpoint}}
\put(477,1154){\usebox{\plotpoint}}
\put(479,1152){\usebox{\plotpoint}}
\put(481,1150){\usebox{\plotpoint}}
\put(483,1147){\usebox{\plotpoint}}
\put(485,1145){\usebox{\plotpoint}}
\put(487,1143){\usebox{\plotpoint}}
\put(489,1140){\usebox{\plotpoint}}
\put(491,1138){\usebox{\plotpoint}}
\put(493,1136){\usebox{\plotpoint}}
\put(495,1133){\usebox{\plotpoint}}
\put(497,1131){\usebox{\plotpoint}}
\put(499,1129){\usebox{\plotpoint}}
\put(501,1126){\usebox{\plotpoint}}
\put(503,1124){\usebox{\plotpoint}}
\put(505,1122){\usebox{\plotpoint}}
\put(507,1119){\usebox{\plotpoint}}
\put(509,1117){\usebox{\plotpoint}}
\put(511,1115){\usebox{\plotpoint}}
\put(513,1112){\usebox{\plotpoint}}
\put(515,1110){\usebox{\plotpoint}}
\put(517,1107){\usebox{\plotpoint}}
\put(519,1105){\usebox{\plotpoint}}
\put(521,1102){\usebox{\plotpoint}}
\put(523,1100){\usebox{\plotpoint}}
\put(525,1097){\usebox{\plotpoint}}
\put(527,1095){\usebox{\plotpoint}}
\put(529,1093){\usebox{\plotpoint}}
\put(531,1090){\usebox{\plotpoint}}
\put(533,1088){\usebox{\plotpoint}}
\put(535,1086){\usebox{\plotpoint}}
\put(537,1083){\usebox{\plotpoint}}
\put(539,1081){\usebox{\plotpoint}}
\put(541,1078){\usebox{\plotpoint}}
\put(543,1076){\usebox{\plotpoint}}
\put(545,1073){\usebox{\plotpoint}}
\put(547,1071){\usebox{\plotpoint}}
\put(549,1068){\usebox{\plotpoint}}
\put(551,1066){\usebox{\plotpoint}}
\put(553,1063){\usebox{\plotpoint}}
\put(555,1061){\usebox{\plotpoint}}
\put(557,1058){\usebox{\plotpoint}}
\put(559,1056){\usebox{\plotpoint}}
\put(561,1053){\usebox{\plotpoint}}
\put(563,1051){\usebox{\plotpoint}}
\put(565,1049){\usebox{\plotpoint}}
\put(567,1046){\usebox{\plotpoint}}
\put(569,1044){\usebox{\plotpoint}}
\put(571,1042){\usebox{\plotpoint}}
\put(573,1039){\usebox{\plotpoint}}
\put(575,1036){\usebox{\plotpoint}}
\put(577,1033){\usebox{\plotpoint}}
\put(579,1031){\usebox{\plotpoint}}
\put(581,1028){\usebox{\plotpoint}}
\put(583,1026){\usebox{\plotpoint}}
\put(585,1023){\usebox{\plotpoint}}
\put(587,1020){\usebox{\plotpoint}}
\put(589,1017){\usebox{\plotpoint}}
\put(591,1015){\usebox{\plotpoint}}
\put(593,1012){\usebox{\plotpoint}}
\put(595,1009){\usebox{\plotpoint}}
\put(597,1007){\usebox{\plotpoint}}
\put(599,1004){\usebox{\plotpoint}}
\put(601,1001){\usebox{\plotpoint}}
\put(603,999){\usebox{\plotpoint}}
\put(605,996){\usebox{\plotpoint}}
\put(607,993){\usebox{\plotpoint}}
\put(609,991){\usebox{\plotpoint}}
\put(611,988){\usebox{\plotpoint}}
\put(613,985){\usebox{\plotpoint}}
\put(615,983){\usebox{\plotpoint}}
\put(617,980){\usebox{\plotpoint}}
\put(619,977){\usebox{\plotpoint}}
\put(621,974){\usebox{\plotpoint}}
\put(623,971){\usebox{\plotpoint}}
\put(625,969){\usebox{\plotpoint}}
\put(627,966){\usebox{\plotpoint}}
\put(629,963){\usebox{\plotpoint}}
\put(631,960){\usebox{\plotpoint}}
\put(633,957){\usebox{\plotpoint}}
\put(635,954){\usebox{\plotpoint}}
\put(637,952){\usebox{\plotpoint}}
\put(639,949){\usebox{\plotpoint}}
\put(641,946){\usebox{\plotpoint}}
\put(643,943){\usebox{\plotpoint}}
\put(645,940){\usebox{\plotpoint}}
\put(647,937){\usebox{\plotpoint}}
\put(649,933){\usebox{\plotpoint}}
\put(651,930){\usebox{\plotpoint}}
\put(653,927){\usebox{\plotpoint}}
\put(655,924){\usebox{\plotpoint}}
\put(657,920){\usebox{\plotpoint}}
\put(659,917){\usebox{\plotpoint}}
\put(661,914){\usebox{\plotpoint}}
\put(663,910){\usebox{\plotpoint}}
\put(665,907){\usebox{\plotpoint}}
\put(667,904){\usebox{\plotpoint}}
\put(669,900){\usebox{\plotpoint}}
\put(671,896){\usebox{\plotpoint}}
\put(673,892){\usebox{\plotpoint}}
\put(675,888){\usebox{\plotpoint}}
\put(677,884){\usebox{\plotpoint}}
\put(679,879){\usebox{\plotpoint}}
\put(681,874){\usebox{\plotpoint}}
\put(683,870){\usebox{\plotpoint}}
\put(685,865){\usebox{\plotpoint}}
\put(687,860){\usebox{\plotpoint}}
\put(689,856){\usebox{\plotpoint}}
\put(691,848){\usebox{\plotpoint}}
\put(692,844){\usebox{\plotpoint}}
\put(693,840){\usebox{\plotpoint}}
\put(694,836){\usebox{\plotpoint}}
\put(695,833){\usebox{\plotpoint}}
\put(696,829){\usebox{\plotpoint}}
\put(697,825){\usebox{\plotpoint}}
\put(698,821){\usebox{\plotpoint}}
\put(699,817){\usebox{\plotpoint}}
\put(700,813){\usebox{\plotpoint}}
\put(701,810){\usebox{\plotpoint}}
\put(702,810){\usebox{\plotpoint}}
\put(998,810){\usebox{\plotpoint}}
\put(998,810){\usebox{\plotpoint}}
\put(999,813){\usebox{\plotpoint}}
\put(1000,817){\usebox{\plotpoint}}
\put(1001,821){\usebox{\plotpoint}}
\put(1002,825){\usebox{\plotpoint}}
\put(1003,829){\usebox{\plotpoint}}
\put(1004,832){\usebox{\plotpoint}}
\put(1005,836){\usebox{\plotpoint}}
\put(1006,840){\usebox{\plotpoint}}
\put(1007,844){\usebox{\plotpoint}}
\put(1008,848){\usebox{\plotpoint}}
\put(1009,852){\usebox{\plotpoint}}
\put(1011,858){\usebox{\plotpoint}}
\put(1013,862){\usebox{\plotpoint}}
\put(1015,867){\usebox{\plotpoint}}
\put(1017,872){\usebox{\plotpoint}}
\put(1019,876){\usebox{\plotpoint}}
\put(1021,881){\usebox{\plotpoint}}
\put(1023,886){\usebox{\plotpoint}}
\put(1025,890){\usebox{\plotpoint}}
\put(1027,894){\usebox{\plotpoint}}
\put(1029,898){\usebox{\plotpoint}}
\put(1031,902){\usebox{\plotpoint}}
\put(1033,906){\usebox{\plotpoint}}
\put(1035,909){\usebox{\plotpoint}}
\put(1037,912){\usebox{\plotpoint}}
\put(1039,916){\usebox{\plotpoint}}
\put(1041,919){\usebox{\plotpoint}}
\put(1043,922){\usebox{\plotpoint}}
\put(1045,926){\usebox{\plotpoint}}
\put(1047,929){\usebox{\plotpoint}}
\put(1049,932){\usebox{\plotpoint}}
\put(1051,935){\usebox{\plotpoint}}
\put(1053,938){\usebox{\plotpoint}}
\put(1055,941){\usebox{\plotpoint}}
\put(1057,944){\usebox{\plotpoint}}
\put(1059,947){\usebox{\plotpoint}}
\put(1061,950){\usebox{\plotpoint}}
\put(1063,953){\usebox{\plotpoint}}
\put(1065,956){\usebox{\plotpoint}}
\put(1067,959){\usebox{\plotpoint}}
\put(1069,962){\usebox{\plotpoint}}
\put(1071,964){\usebox{\plotpoint}}
\put(1073,967){\usebox{\plotpoint}}
\put(1075,970){\usebox{\plotpoint}}
\put(1077,973){\usebox{\plotpoint}}
\put(1079,976){\usebox{\plotpoint}}
\put(1081,979){\usebox{\plotpoint}}
\put(1083,981){\usebox{\plotpoint}}
\put(1085,984){\usebox{\plotpoint}}
\put(1087,986){\usebox{\plotpoint}}
\put(1089,989){\usebox{\plotpoint}}
\put(1091,992){\usebox{\plotpoint}}
\put(1093,994){\usebox{\plotpoint}}
\put(1095,997){\usebox{\plotpoint}}
\put(1097,1000){\usebox{\plotpoint}}
\put(1099,1002){\usebox{\plotpoint}}
\put(1101,1005){\usebox{\plotpoint}}
\put(1103,1008){\usebox{\plotpoint}}
\put(1105,1010){\usebox{\plotpoint}}
\put(1107,1013){\usebox{\plotpoint}}
\put(1109,1016){\usebox{\plotpoint}}
\put(1111,1019){\usebox{\plotpoint}}
\put(1113,1021){\usebox{\plotpoint}}
\put(1115,1024){\usebox{\plotpoint}}
\put(1117,1027){\usebox{\plotpoint}}
\put(1119,1030){\usebox{\plotpoint}}
\put(1121,1032){\usebox{\plotpoint}}
\put(1123,1035){\usebox{\plotpoint}}
\put(1125,1038){\usebox{\plotpoint}}
\put(1127,1040){\usebox{\plotpoint}}
\put(1129,1043){\usebox{\plotpoint}}
\put(1131,1045){\usebox{\plotpoint}}
\put(1133,1047){\usebox{\plotpoint}}
\put(1135,1050){\usebox{\plotpoint}}
\put(1137,1052){\usebox{\plotpoint}}
\put(1139,1054){\usebox{\plotpoint}}
\put(1141,1057){\usebox{\plotpoint}}
\put(1143,1059){\usebox{\plotpoint}}
\put(1145,1062){\usebox{\plotpoint}}
\put(1147,1064){\usebox{\plotpoint}}
\put(1149,1067){\usebox{\plotpoint}}
\put(1151,1069){\usebox{\plotpoint}}
\put(1153,1072){\usebox{\plotpoint}}
\put(1155,1074){\usebox{\plotpoint}}
\put(1157,1077){\usebox{\plotpoint}}
\put(1159,1079){\usebox{\plotpoint}}
\put(1161,1082){\usebox{\plotpoint}}
\put(1163,1084){\usebox{\plotpoint}}
\put(1165,1087){\usebox{\plotpoint}}
\put(1167,1089){\usebox{\plotpoint}}
\put(1169,1091){\usebox{\plotpoint}}
\put(1171,1094){\usebox{\plotpoint}}
\put(1173,1096){\usebox{\plotpoint}}
\put(1175,1098){\usebox{\plotpoint}}
\put(1177,1101){\usebox{\plotpoint}}
\put(1179,1103){\usebox{\plotpoint}}
\put(1181,1106){\usebox{\plotpoint}}
\put(1183,1108){\usebox{\plotpoint}}
\put(1185,1111){\usebox{\plotpoint}}
\put(1187,1113){\usebox{\plotpoint}}
\put(1189,1116){\usebox{\plotpoint}}
\put(1191,1118){\usebox{\plotpoint}}
\put(1193,1120){\usebox{\plotpoint}}
\put(1195,1123){\usebox{\plotpoint}}
\put(1197,1125){\usebox{\plotpoint}}
\put(1199,1127){\usebox{\plotpoint}}
\put(1201,1130){\usebox{\plotpoint}}
\put(1203,1132){\usebox{\plotpoint}}
\put(1205,1134){\usebox{\plotpoint}}
\put(1207,1137){\usebox{\plotpoint}}
\put(1209,1139){\usebox{\plotpoint}}
\put(1211,1141){\usebox{\plotpoint}}
\put(1213,1144){\usebox{\plotpoint}}
\put(1215,1146){\usebox{\plotpoint}}
\put(1217,1148){\usebox{\plotpoint}}
\put(1219,1151){\usebox{\plotpoint}}
\put(1221,1153){\usebox{\plotpoint}}
\put(1223,1155){\usebox{\plotpoint}}
\put(1225,1158){\usebox{\plotpoint}}
\put(1227,1160){\usebox{\plotpoint}}
\put(1229,1162){\usebox{\plotpoint}}
\put(1231,1165){\usebox{\plotpoint}}
\put(1233,1167){\usebox{\plotpoint}}
\put(1235,1169){\usebox{\plotpoint}}
\put(1237,1172){\usebox{\plotpoint}}
\put(1239,1174){\usebox{\plotpoint}}
\put(1241,1176){\usebox{\plotpoint}}
\put(1243,1178){\usebox{\plotpoint}}
\put(1245,1180){\usebox{\plotpoint}}
\put(1247,1183){\usebox{\plotpoint}}
\put(1249,1185){\usebox{\plotpoint}}
\put(1251,1188){\usebox{\plotpoint}}
\put(1253,1190){\usebox{\plotpoint}}
\put(1255,1193){\usebox{\plotpoint}}
\put(1257,1195){\usebox{\plotpoint}}
\put(1259,1198){\usebox{\plotpoint}}
\put(1261,1200){\usebox{\plotpoint}}
\put(1263,1202){\usebox{\plotpoint}}
\put(1265,1205){\usebox{\plotpoint}}
\put(1267,1207){\usebox{\plotpoint}}
\put(1269,1209){\usebox{\plotpoint}}
\put(1271,1212){\usebox{\plotpoint}}
\put(1273,1214){\usebox{\plotpoint}}
\put(1275,1216){\usebox{\plotpoint}}
\put(1277,1218){\usebox{\plotpoint}}
\put(1279,1220){\usebox{\plotpoint}}
\put(1281,1222){\usebox{\plotpoint}}
\put(1283,1225){\usebox{\plotpoint}}
\put(1285,1227){\usebox{\plotpoint}}
\put(1287,1229){\usebox{\plotpoint}}
\put(1289,1232){\usebox{\plotpoint}}
\put(1291,1234){\usebox{\plotpoint}}
\put(1293,1236){\usebox{\plotpoint}}
\put(1295,1239){\usebox{\plotpoint}}
\put(1297,1241){\usebox{\plotpoint}}
\put(1299,1243){\usebox{\plotpoint}}
\put(1301,1245){\usebox{\plotpoint}}
\put(1303,1247){\usebox{\plotpoint}}
\put(1305,1249){\usebox{\plotpoint}}
\put(1307,1252){\usebox{\plotpoint}}
\put(1309,1254){\usebox{\plotpoint}}
\put(1311,1256){\usebox{\plotpoint}}
\put(1313,1259){\usebox{\plotpoint}}
\put(1315,1261){\usebox{\plotpoint}}
\put(1317,1263){\usebox{\plotpoint}}
\put(1319,1266){\usebox{\plotpoint}}
\put(1321,1268){\usebox{\plotpoint}}
\put(1323,1270){\usebox{\plotpoint}}
\put(1325,1273){\usebox{\plotpoint}}
\put(1327,1275){\usebox{\plotpoint}}
\put(1329,1277){\usebox{\plotpoint}}
\put(1331,1280){\usebox{\plotpoint}}
\put(1333,1282){\usebox{\plotpoint}}
\put(1335,1284){\usebox{\plotpoint}}
\put(1337,1286){\usebox{\plotpoint}}
\put(1339,1288){\usebox{\plotpoint}}
\put(1341,1291){\usebox{\plotpoint}}
\put(1343,1293){\usebox{\plotpoint}}
\put(1345,1295){\usebox{\plotpoint}}
\put(1347,1297){\usebox{\plotpoint}}
\put(1349,1300){\usebox{\plotpoint}}
\put(1351,1302){\usebox{\plotpoint}}
\put(1353,1304){\usebox{\plotpoint}}
\put(1355,1307){\usebox{\plotpoint}}
\put(1357,1309){\usebox{\plotpoint}}
\put(1359,1311){\usebox{\plotpoint}}
\put(1361,1313){\usebox{\plotpoint}}
\put(1363,1315){\usebox{\plotpoint}}
\put(1365,1318){\usebox{\plotpoint}}
\put(1367,1320){\usebox{\plotpoint}}
\put(1369,1322){\usebox{\plotpoint}}
\put(1371,1324){\usebox{\plotpoint}}
\put(1373,1326){\usebox{\plotpoint}}
\put(1375,1328){\usebox{\plotpoint}}
\put(1377,1331){\usebox{\plotpoint}}
\put(1379,1333){\usebox{\plotpoint}}
\put(1381,1335){\usebox{\plotpoint}}
\put(1383,1337){\usebox{\plotpoint}}
\put(1385,1339){\usebox{\plotpoint}}
\put(1387,1341){\usebox{\plotpoint}}
\put(1389,1344){\usebox{\plotpoint}}
\put(1391,1346){\usebox{\plotpoint}}
\put(1393,1349){\usebox{\plotpoint}}
\put(1395,1351){\usebox{\plotpoint}}
\put(1397,1354){\usebox{\plotpoint}}
\put(1399,1356){\usebox{\plotpoint}}
\put(1401,1359){\usebox{\plotpoint}}
\put(1403,1361){\usebox{\plotpoint}}
\put(1405,1363){\usebox{\plotpoint}}
\put(1407,1365){\usebox{\plotpoint}}
\put(1409,1367){\usebox{\plotpoint}}
\put(1411,1369){\usebox{\plotpoint}}
\put(1413,1372){\usebox{\plotpoint}}
\put(1415,1374){\usebox{\plotpoint}}
\put(1417,1376){\usebox{\plotpoint}}
\put(1419,1378){\usebox{\plotpoint}}
\put(1421,1380){\usebox{\plotpoint}}
\put(1423,1382){\usebox{\plotpoint}}
\put(1425,1385){\usebox{\plotpoint}}
\put(1427,1387){\usebox{\plotpoint}}
\put(1429,1389){\usebox{\plotpoint}}
\put(1431,1391){\usebox{\plotpoint}}
\put(1433,1393){\usebox{\plotpoint}}
\put(1435,1395){\usebox{\plotpoint}}
\put(264,178){\usebox{\plotpoint}}
\put(266,180){\usebox{\plotpoint}}
\put(268,182){\usebox{\plotpoint}}
\put(270,184){\usebox{\plotpoint}}
\put(272,186){\usebox{\plotpoint}}
\put(274,188){\usebox{\plotpoint}}
\put(276,190){\usebox{\plotpoint}}
\put(278,193){\usebox{\plotpoint}}
\put(280,195){\usebox{\plotpoint}}
\put(282,197){\usebox{\plotpoint}}
\put(284,199){\usebox{\plotpoint}}
\put(286,201){\usebox{\plotpoint}}
\put(288,203){\usebox{\plotpoint}}
\put(290,206){\usebox{\plotpoint}}
\put(292,208){\usebox{\plotpoint}}
\put(294,210){\usebox{\plotpoint}}
\put(296,212){\usebox{\plotpoint}}
\put(298,214){\usebox{\plotpoint}}
\put(300,216){\usebox{\plotpoint}}
\put(302,219){\usebox{\plotpoint}}
\put(304,222){\usebox{\plotpoint}}
\put(306,224){\usebox{\plotpoint}}
\put(308,227){\usebox{\plotpoint}}
\put(310,229){\usebox{\plotpoint}}
\put(312,232){\usebox{\plotpoint}}
\put(314,234){\usebox{\plotpoint}}
\put(316,236){\usebox{\plotpoint}}
\put(318,238){\usebox{\plotpoint}}
\put(320,240){\usebox{\plotpoint}}
\put(322,242){\usebox{\plotpoint}}
\put(324,245){\usebox{\plotpoint}}
\put(326,247){\usebox{\plotpoint}}
\put(328,249){\usebox{\plotpoint}}
\put(330,251){\usebox{\plotpoint}}
\put(332,253){\usebox{\plotpoint}}
\put(334,255){\usebox{\plotpoint}}
\put(336,258){\usebox{\plotpoint}}
\put(338,260){\usebox{\plotpoint}}
\put(340,262){\usebox{\plotpoint}}
\put(342,264){\usebox{\plotpoint}}
\put(344,266){\usebox{\plotpoint}}
\put(346,268){\usebox{\plotpoint}}
\put(348,271){\usebox{\plotpoint}}
\put(350,273){\usebox{\plotpoint}}
\put(352,275){\usebox{\plotpoint}}
\put(354,278){\usebox{\plotpoint}}
\put(356,280){\usebox{\plotpoint}}
\put(358,282){\usebox{\plotpoint}}
\put(360,285){\usebox{\plotpoint}}
\put(362,287){\usebox{\plotpoint}}
\put(364,289){\usebox{\plotpoint}}
\put(366,291){\usebox{\plotpoint}}
\put(368,293){\usebox{\plotpoint}}
\put(370,295){\usebox{\plotpoint}}
\put(372,298){\usebox{\plotpoint}}
\put(374,300){\usebox{\plotpoint}}
\put(376,302){\usebox{\plotpoint}}
\put(378,305){\usebox{\plotpoint}}
\put(380,307){\usebox{\plotpoint}}
\put(382,310){\usebox{\plotpoint}}
\put(384,312){\usebox{\plotpoint}}
\put(386,314){\usebox{\plotpoint}}
\put(388,316){\usebox{\plotpoint}}
\put(390,319){\usebox{\plotpoint}}
\put(392,321){\usebox{\plotpoint}}
\put(394,323){\usebox{\plotpoint}}
\put(396,326){\usebox{\plotpoint}}
\put(398,328){\usebox{\plotpoint}}
\put(400,330){\usebox{\plotpoint}}
\put(402,332){\usebox{\plotpoint}}
\put(404,334){\usebox{\plotpoint}}
\put(406,337){\usebox{\plotpoint}}
\put(408,339){\usebox{\plotpoint}}
\put(410,341){\usebox{\plotpoint}}
\put(412,343){\usebox{\plotpoint}}
\put(414,346){\usebox{\plotpoint}}
\put(416,348){\usebox{\plotpoint}}
\put(418,350){\usebox{\plotpoint}}
\put(420,353){\usebox{\plotpoint}}
\put(422,355){\usebox{\plotpoint}}
\put(424,357){\usebox{\plotpoint}}
\put(426,359){\usebox{\plotpoint}}
\put(428,361){\usebox{\plotpoint}}
\put(430,364){\usebox{\plotpoint}}
\put(432,366){\usebox{\plotpoint}}
\put(434,368){\usebox{\plotpoint}}
\put(436,370){\usebox{\plotpoint}}
\put(438,373){\usebox{\plotpoint}}
\put(440,375){\usebox{\plotpoint}}
\put(442,377){\usebox{\plotpoint}}
\put(444,380){\usebox{\plotpoint}}
\put(446,383){\usebox{\plotpoint}}
\put(448,385){\usebox{\plotpoint}}
\put(450,388){\usebox{\plotpoint}}
\put(452,390){\usebox{\plotpoint}}
\put(454,393){\usebox{\plotpoint}}
\put(456,395){\usebox{\plotpoint}}
\put(458,397){\usebox{\plotpoint}}
\put(460,399){\usebox{\plotpoint}}
\put(462,401){\usebox{\plotpoint}}
\put(464,403){\usebox{\plotpoint}}
\put(466,406){\usebox{\plotpoint}}
\put(468,408){\usebox{\plotpoint}}
\put(470,410){\usebox{\plotpoint}}
\put(472,413){\usebox{\plotpoint}}
\put(474,415){\usebox{\plotpoint}}
\put(476,417){\usebox{\plotpoint}}
\put(478,420){\usebox{\plotpoint}}
\put(480,422){\usebox{\plotpoint}}
\put(482,424){\usebox{\plotpoint}}
\put(484,427){\usebox{\plotpoint}}
\put(486,429){\usebox{\plotpoint}}
\put(488,431){\usebox{\plotpoint}}
\put(490,434){\usebox{\plotpoint}}
\put(492,436){\usebox{\plotpoint}}
\put(494,438){\usebox{\plotpoint}}
\put(496,441){\usebox{\plotpoint}}
\put(498,443){\usebox{\plotpoint}}
\put(500,445){\usebox{\plotpoint}}
\put(502,448){\usebox{\plotpoint}}
\put(504,450){\usebox{\plotpoint}}
\put(506,452){\usebox{\plotpoint}}
\put(508,455){\usebox{\plotpoint}}
\put(510,457){\usebox{\plotpoint}}
\put(512,459){\usebox{\plotpoint}}
\put(514,462){\usebox{\plotpoint}}
\put(516,464){\usebox{\plotpoint}}
\put(518,467){\usebox{\plotpoint}}
\put(520,469){\usebox{\plotpoint}}
\put(522,472){\usebox{\plotpoint}}
\put(524,475){\usebox{\plotpoint}}
\put(526,477){\usebox{\plotpoint}}
\put(528,479){\usebox{\plotpoint}}
\put(530,481){\usebox{\plotpoint}}
\put(532,484){\usebox{\plotpoint}}
\put(534,486){\usebox{\plotpoint}}
\put(536,488){\usebox{\plotpoint}}
\put(538,491){\usebox{\plotpoint}}
\put(540,494){\usebox{\plotpoint}}
\put(542,496){\usebox{\plotpoint}}
\put(544,499){\usebox{\plotpoint}}
\put(546,501){\usebox{\plotpoint}}
\put(548,504){\usebox{\plotpoint}}
\put(550,506){\usebox{\plotpoint}}
\put(552,509){\usebox{\plotpoint}}
\put(554,511){\usebox{\plotpoint}}
\put(556,514){\usebox{\plotpoint}}
\put(558,516){\usebox{\plotpoint}}
\put(560,519){\usebox{\plotpoint}}
\put(562,521){\usebox{\plotpoint}}
\put(564,523){\usebox{\plotpoint}}
\put(566,526){\usebox{\plotpoint}}
\put(568,528){\usebox{\plotpoint}}
\put(570,530){\usebox{\plotpoint}}
\put(572,533){\usebox{\plotpoint}}
\put(574,535){\usebox{\plotpoint}}
\put(576,538){\usebox{\plotpoint}}
\put(578,540){\usebox{\plotpoint}}
\put(580,543){\usebox{\plotpoint}}
\put(582,546){\usebox{\plotpoint}}
\put(584,548){\usebox{\plotpoint}}
\put(586,551){\usebox{\plotpoint}}
\put(588,554){\usebox{\plotpoint}}
\put(590,557){\usebox{\plotpoint}}
\put(592,559){\usebox{\plotpoint}}
\put(594,562){\usebox{\plotpoint}}
\put(596,565){\usebox{\plotpoint}}
\put(598,567){\usebox{\plotpoint}}
\put(600,570){\usebox{\plotpoint}}
\put(602,573){\usebox{\plotpoint}}
\put(604,575){\usebox{\plotpoint}}
\put(606,578){\usebox{\plotpoint}}
\put(608,581){\usebox{\plotpoint}}
\put(610,583){\usebox{\plotpoint}}
\put(612,586){\usebox{\plotpoint}}
\put(614,589){\usebox{\plotpoint}}
\put(616,591){\usebox{\plotpoint}}
\put(618,594){\usebox{\plotpoint}}
\put(620,597){\usebox{\plotpoint}}
\put(622,600){\usebox{\plotpoint}}
\put(624,603){\usebox{\plotpoint}}
\put(626,605){\usebox{\plotpoint}}
\put(628,608){\usebox{\plotpoint}}
\put(630,611){\usebox{\plotpoint}}
\put(632,614){\usebox{\plotpoint}}
\put(634,617){\usebox{\plotpoint}}
\put(636,620){\usebox{\plotpoint}}
\put(638,622){\usebox{\plotpoint}}
\put(640,625){\usebox{\plotpoint}}
\put(642,628){\usebox{\plotpoint}}
\put(644,631){\usebox{\plotpoint}}
\put(646,634){\usebox{\plotpoint}}
\put(648,637){\usebox{\plotpoint}}
\put(650,641){\usebox{\plotpoint}}
\put(652,644){\usebox{\plotpoint}}
\put(654,647){\usebox{\plotpoint}}
\put(656,650){\usebox{\plotpoint}}
\put(657,652){\usebox{\plotpoint}}
\put(658,654){\usebox{\plotpoint}}
\put(659,655){\usebox{\plotpoint}}
\put(660,657){\usebox{\plotpoint}}
\put(661,659){\usebox{\plotpoint}}
\put(662,660){\usebox{\plotpoint}}
\put(663,662){\usebox{\plotpoint}}
\put(664,664){\usebox{\plotpoint}}
\put(665,665){\usebox{\plotpoint}}
\put(666,667){\usebox{\plotpoint}}
\put(667,669){\usebox{\plotpoint}}
\put(668,671){\usebox{\plotpoint}}
\put(669,673){\usebox{\plotpoint}}
\put(670,675){\usebox{\plotpoint}}
\put(671,677){\usebox{\plotpoint}}
\put(672,679){\usebox{\plotpoint}}
\put(673,681){\usebox{\plotpoint}}
\put(674,683){\usebox{\plotpoint}}
\put(675,685){\usebox{\plotpoint}}
\put(676,687){\usebox{\plotpoint}}
\put(677,689){\usebox{\plotpoint}}
\put(678,691){\usebox{\plotpoint}}
\put(679,693){\usebox{\plotpoint}}
\put(680,695){\usebox{\plotpoint}}
\put(681,697){\usebox{\plotpoint}}
\put(682,700){\usebox{\plotpoint}}
\put(683,702){\usebox{\plotpoint}}
\put(684,704){\usebox{\plotpoint}}
\put(685,707){\usebox{\plotpoint}}
\put(686,709){\usebox{\plotpoint}}
\put(687,711){\usebox{\plotpoint}}
\put(688,714){\usebox{\plotpoint}}
\put(689,716){\usebox{\plotpoint}}
\put(690,718){\usebox{\plotpoint}}
\put(691,722){\usebox{\plotpoint}}
\put(692,726){\usebox{\plotpoint}}
\put(693,730){\usebox{\plotpoint}}
\put(694,734){\usebox{\plotpoint}}
\put(695,738){\usebox{\plotpoint}}
\put(696,741){\usebox{\plotpoint}}
\put(697,745){\usebox{\plotpoint}}
\put(698,749){\usebox{\plotpoint}}
\put(699,753){\usebox{\plotpoint}}
\put(700,757){\usebox{\plotpoint}}
\put(701,761){\usebox{\plotpoint}}
\put(702,764){\usebox{\plotpoint}}
\put(998,765){\usebox{\plotpoint}}
\put(998,761){\usebox{\plotpoint}}
\put(999,757){\usebox{\plotpoint}}
\put(1000,753){\usebox{\plotpoint}}
\put(1001,749){\usebox{\plotpoint}}
\put(1002,745){\usebox{\plotpoint}}
\put(1003,742){\usebox{\plotpoint}}
\put(1004,738){\usebox{\plotpoint}}
\put(1005,734){\usebox{\plotpoint}}
\put(1006,730){\usebox{\plotpoint}}
\put(1007,726){\usebox{\plotpoint}}
\put(1008,722){\usebox{\plotpoint}}
\put(1009,719){\usebox{\plotpoint}}
\put(1010,716){\usebox{\plotpoint}}
\put(1011,714){\usebox{\plotpoint}}
\put(1012,712){\usebox{\plotpoint}}
\put(1013,709){\usebox{\plotpoint}}
\put(1014,707){\usebox{\plotpoint}}
\put(1015,705){\usebox{\plotpoint}}
\put(1016,702){\usebox{\plotpoint}}
\put(1017,700){\usebox{\plotpoint}}
\put(1018,698){\usebox{\plotpoint}}
\put(1019,695){\usebox{\plotpoint}}
\put(1020,693){\usebox{\plotpoint}}
\put(1021,691){\usebox{\plotpoint}}
\put(1022,689){\usebox{\plotpoint}}
\put(1023,687){\usebox{\plotpoint}}
\put(1024,685){\usebox{\plotpoint}}
\put(1025,683){\usebox{\plotpoint}}
\put(1026,681){\usebox{\plotpoint}}
\put(1027,679){\usebox{\plotpoint}}
\put(1028,677){\usebox{\plotpoint}}
\put(1029,675){\usebox{\plotpoint}}
\put(1030,673){\usebox{\plotpoint}}
\put(1031,671){\usebox{\plotpoint}}
\put(1032,669){\usebox{\plotpoint}}
\put(1033,667){\usebox{\plotpoint}}
\put(1034,665){\usebox{\plotpoint}}
\put(1035,663){\usebox{\plotpoint}}
\put(1036,662){\usebox{\plotpoint}}
\put(1037,660){\usebox{\plotpoint}}
\put(1038,658){\usebox{\plotpoint}}
\put(1039,657){\usebox{\plotpoint}}
\put(1040,655){\usebox{\plotpoint}}
\put(1041,653){\usebox{\plotpoint}}
\put(1042,652){\usebox{\plotpoint}}
\put(1044,649){\usebox{\plotpoint}}
\put(1046,645){\usebox{\plotpoint}}
\put(1048,642){\usebox{\plotpoint}}
\put(1050,639){\usebox{\plotpoint}}
\put(1052,636){\usebox{\plotpoint}}
\put(1054,633){\usebox{\plotpoint}}
\put(1056,630){\usebox{\plotpoint}}
\put(1058,627){\usebox{\plotpoint}}
\put(1060,624){\usebox{\plotpoint}}
\put(1062,621){\usebox{\plotpoint}}
\put(1064,618){\usebox{\plotpoint}}
\put(1066,615){\usebox{\plotpoint}}
\put(1068,613){\usebox{\plotpoint}}
\put(1070,610){\usebox{\plotpoint}}
\put(1072,607){\usebox{\plotpoint}}
\put(1074,604){\usebox{\plotpoint}}
\put(1076,601){\usebox{\plotpoint}}
\put(1078,598){\usebox{\plotpoint}}
\put(1080,596){\usebox{\plotpoint}}
\put(1082,593){\usebox{\plotpoint}}
\put(1084,590){\usebox{\plotpoint}}
\put(1086,588){\usebox{\plotpoint}}
\put(1088,585){\usebox{\plotpoint}}
\put(1090,582){\usebox{\plotpoint}}
\put(1092,580){\usebox{\plotpoint}}
\put(1094,577){\usebox{\plotpoint}}
\put(1096,574){\usebox{\plotpoint}}
\put(1098,572){\usebox{\plotpoint}}
\put(1100,569){\usebox{\plotpoint}}
\put(1102,566){\usebox{\plotpoint}}
\put(1104,564){\usebox{\plotpoint}}
\put(1106,561){\usebox{\plotpoint}}
\put(1108,558){\usebox{\plotpoint}}
\put(1110,555){\usebox{\plotpoint}}
\put(1112,553){\usebox{\plotpoint}}
\put(1114,550){\usebox{\plotpoint}}
\put(1116,547){\usebox{\plotpoint}}
\put(1118,545){\usebox{\plotpoint}}
\put(1120,542){\usebox{\plotpoint}}
\put(1122,539){\usebox{\plotpoint}}
\put(1124,537){\usebox{\plotpoint}}
\put(1126,534){\usebox{\plotpoint}}
\put(1128,531){\usebox{\plotpoint}}
\put(1130,529){\usebox{\plotpoint}}
\put(1132,527){\usebox{\plotpoint}}
\put(1134,524){\usebox{\plotpoint}}
\put(1136,522){\usebox{\plotpoint}}
\put(1138,520){\usebox{\plotpoint}}
\put(1140,517){\usebox{\plotpoint}}
\put(1142,515){\usebox{\plotpoint}}
\put(1144,512){\usebox{\plotpoint}}
\put(1146,510){\usebox{\plotpoint}}
\put(1148,507){\usebox{\plotpoint}}
\put(1150,505){\usebox{\plotpoint}}
\put(1152,502){\usebox{\plotpoint}}
\put(1154,500){\usebox{\plotpoint}}
\put(1156,497){\usebox{\plotpoint}}
\put(1158,495){\usebox{\plotpoint}}
\put(1160,492){\usebox{\plotpoint}}
\put(1162,490){\usebox{\plotpoint}}
\put(1164,487){\usebox{\plotpoint}}
\put(1166,485){\usebox{\plotpoint}}
\put(1168,483){\usebox{\plotpoint}}
\put(1170,480){\usebox{\plotpoint}}
\put(1172,478){\usebox{\plotpoint}}
\put(1174,476){\usebox{\plotpoint}}
\put(1176,473){\usebox{\plotpoint}}
\put(1178,471){\usebox{\plotpoint}}
\put(1180,468){\usebox{\plotpoint}}
\put(1182,466){\usebox{\plotpoint}}
\put(1184,463){\usebox{\plotpoint}}
\put(1186,461){\usebox{\plotpoint}}
\put(1188,458){\usebox{\plotpoint}}
\put(1190,456){\usebox{\plotpoint}}
\put(1192,454){\usebox{\plotpoint}}
\put(1194,451){\usebox{\plotpoint}}
\put(1196,449){\usebox{\plotpoint}}
\put(1198,447){\usebox{\plotpoint}}
\put(1200,444){\usebox{\plotpoint}}
\put(1202,442){\usebox{\plotpoint}}
\put(1204,440){\usebox{\plotpoint}}
\put(1206,437){\usebox{\plotpoint}}
\put(1208,435){\usebox{\plotpoint}}
\put(1210,433){\usebox{\plotpoint}}
\put(1212,430){\usebox{\plotpoint}}
\put(1214,428){\usebox{\plotpoint}}
\put(1216,426){\usebox{\plotpoint}}
\put(1218,423){\usebox{\plotpoint}}
\put(1220,421){\usebox{\plotpoint}}
\put(1222,419){\usebox{\plotpoint}}
\put(1224,416){\usebox{\plotpoint}}
\put(1226,414){\usebox{\plotpoint}}
\put(1228,412){\usebox{\plotpoint}}
\put(1230,409){\usebox{\plotpoint}}
\put(1232,407){\usebox{\plotpoint}}
\put(1234,405){\usebox{\plotpoint}}
\put(1236,402){\usebox{\plotpoint}}
\put(1238,400){\usebox{\plotpoint}}
\put(1240,398){\usebox{\plotpoint}}
\put(1242,396){\usebox{\plotpoint}}
\put(1244,394){\usebox{\plotpoint}}
\put(1246,392){\usebox{\plotpoint}}
\put(1248,389){\usebox{\plotpoint}}
\put(1250,386){\usebox{\plotpoint}}
\put(1252,384){\usebox{\plotpoint}}
\put(1254,381){\usebox{\plotpoint}}
\put(1256,379){\usebox{\plotpoint}}
\put(1258,376){\usebox{\plotpoint}}
\put(1260,374){\usebox{\plotpoint}}
\put(1262,372){\usebox{\plotpoint}}
\put(1264,369){\usebox{\plotpoint}}
\put(1266,367){\usebox{\plotpoint}}
\put(1268,365){\usebox{\plotpoint}}
\put(1270,362){\usebox{\plotpoint}}
\put(1272,360){\usebox{\plotpoint}}
\put(1274,358){\usebox{\plotpoint}}
\put(1276,356){\usebox{\plotpoint}}
\put(1278,354){\usebox{\plotpoint}}
\put(1280,352){\usebox{\plotpoint}}
\put(1282,349){\usebox{\plotpoint}}
\put(1284,347){\usebox{\plotpoint}}
\put(1286,345){\usebox{\plotpoint}}
\put(1288,342){\usebox{\plotpoint}}
\put(1290,340){\usebox{\plotpoint}}
\put(1292,338){\usebox{\plotpoint}}
\put(1294,335){\usebox{\plotpoint}}
\put(1296,333){\usebox{\plotpoint}}
\put(1298,331){\usebox{\plotpoint}}
\put(1300,329){\usebox{\plotpoint}}
\put(1302,327){\usebox{\plotpoint}}
\put(1304,325){\usebox{\plotpoint}}
\put(1306,322){\usebox{\plotpoint}}
\put(1308,320){\usebox{\plotpoint}}
\put(1310,318){\usebox{\plotpoint}}
\put(1312,315){\usebox{\plotpoint}}
\put(1314,313){\usebox{\plotpoint}}
\put(1316,311){\usebox{\plotpoint}}
\put(1318,308){\usebox{\plotpoint}}
\put(1320,306){\usebox{\plotpoint}}
\put(1322,304){\usebox{\plotpoint}}
\put(1324,301){\usebox{\plotpoint}}
\put(1326,299){\usebox{\plotpoint}}
\put(1328,297){\usebox{\plotpoint}}
\put(1330,294){\usebox{\plotpoint}}
\put(1332,292){\usebox{\plotpoint}}
\put(1334,290){\usebox{\plotpoint}}
\put(1336,288){\usebox{\plotpoint}}
\put(1338,286){\usebox{\plotpoint}}
\put(1340,284){\usebox{\plotpoint}}
\put(1342,281){\usebox{\plotpoint}}
\put(1344,279){\usebox{\plotpoint}}
\put(1346,277){\usebox{\plotpoint}}
\put(1348,274){\usebox{\plotpoint}}
\put(1350,272){\usebox{\plotpoint}}
\put(1352,270){\usebox{\plotpoint}}
\put(1354,267){\usebox{\plotpoint}}
\put(1356,265){\usebox{\plotpoint}}
\put(1358,263){\usebox{\plotpoint}}
\put(1360,261){\usebox{\plotpoint}}
\put(1362,259){\usebox{\plotpoint}}
\put(1364,257){\usebox{\plotpoint}}
\put(1366,254){\usebox{\plotpoint}}
\put(1368,252){\usebox{\plotpoint}}
\put(1370,250){\usebox{\plotpoint}}
\put(1372,248){\usebox{\plotpoint}}
\put(1374,246){\usebox{\plotpoint}}
\put(1376,244){\usebox{\plotpoint}}
\put(1378,241){\usebox{\plotpoint}}
\put(1380,239){\usebox{\plotpoint}}
\put(1382,237){\usebox{\plotpoint}}
\put(1384,235){\usebox{\plotpoint}}
\put(1386,233){\usebox{\plotpoint}}
\put(1388,231){\usebox{\plotpoint}}
\put(1390,228){\usebox{\plotpoint}}
\put(1392,225){\usebox{\plotpoint}}
\put(1394,223){\usebox{\plotpoint}}
\put(1396,220){\usebox{\plotpoint}}
\put(1398,218){\usebox{\plotpoint}}
\put(1400,215){\usebox{\plotpoint}}
\put(1402,213){\usebox{\plotpoint}}
\put(1404,211){\usebox{\plotpoint}}
\put(1406,209){\usebox{\plotpoint}}
\put(1408,207){\usebox{\plotpoint}}
\put(1410,205){\usebox{\plotpoint}}
\put(1412,202){\usebox{\plotpoint}}
\put(1414,200){\usebox{\plotpoint}}
\put(1416,198){\usebox{\plotpoint}}
\put(1418,196){\usebox{\plotpoint}}
\put(1420,194){\usebox{\plotpoint}}
\put(1422,192){\usebox{\plotpoint}}
\put(1424,189){\usebox{\plotpoint}}
\put(1426,187){\usebox{\plotpoint}}
\put(1428,185){\usebox{\plotpoint}}
\put(1430,183){\usebox{\plotpoint}}
\put(1432,181){\usebox{\plotpoint}}
\put(1434,179){\usebox{\plotpoint}}
\put(1436,178){\usebox{\plotpoint}}
%
%
\put(267,1412){\line(1,-1){60}}
\put(283,1397){\line(1,-1){60}}
\put(299,1379){\line(1,-1){60}}
\put(315,1360){\line(1,-1){60}}
\put(331,1345){\line(1,-1){60}}
\put(347,1326){\line(1,-1){60}}
\put(363,1311){\line(1,-1){60}}
\put(379,1292){\line(1,-1){60}}
\put(395,1274){\line(1,-1){60}}
\put(411,1259){\line(1,-1){60}}
\put(427,1241){\line(1,-1){60}}
\put(443,1223){\line(1,-1){60}}
\put(459,1206){\line(1,-1){60}}
\put(475,1189){\line(1,-1){60}}
\put(491,1173){\line(1,-1){60}}
\put(507,1155){\line(1,-1){60}}
\put(523,1137){\line(1,-1){60}}
\put(538,1121){\line(1,-1){60}}
\put(554,1104){\line(1,-1){60}}
\put(570,1087){\line(1,-1){60}}
\put(586,1070){\line(1,-1){60}}
\put(602,1052){\line(1,-1){60}}
\put(618,1035){\line(1,-1){60}}
\put(634,1018){\line(1,-1){60}}
\put(650,1001){\line(1,-1){60}}
\put(665,987){\line(1,-1){60}}
\put(681,967){\line(1,-1){60}}
\put(697,950){\line(1,-1){60}}
\put(712,936){\line(1,-1){60}}
\put(728,917){\line(1,-1){60}}
\put(744,900){\line(1,-1){60}}
\put(760,883){\line(1,-1){60}}
\put(776,865){\line(1,-1){60}}
\put(791,851){\line(1,-1){60}}
\put(807,833){\line(1,-1){60}}
\put(823,816){\line(1,-1){60}}
\put(839,798){\line(1,-1){60}}
\put(855,781){\line(1,-1){60}}
\put(870,766){\line(1,-1){60}}
\put(886,747){\line(1,-1){60}}
\put(902,730){\line(1,-1){60}}
\put(917,714){\line(1,-1){60}}
\put(933,697){\line(1,-1){60}}
\put(949,680){\line(1,-1){60}}
\put(965,662){\line(1,-1){60}}
\put(981,645){\line(1,-1){60}}
\put(996,631){\line(1,-1){60}}
\put(1012,612){\line(1,-1){60}}
\put(1028,594){\line(1,-1){60}}
\put(1044,579){\line(1,-1){60}}
\put(1060,560){\line(1,-1){60}}
\put(1075,546){\line(1,-1){60}}
\put(1091,528){\line(1,-1){60}}
\put(1107,510){\line(1,-1){60}}
\put(1123,494){\line(1,-1){60}}
\put(1139,476){\line(1,-1){60}}
\put(1155,458){\line(1,-1){60}}
\put(1171,443){\line(1,-1){60}}
\put(1187,423){\line(1,-1){60}}
\put(1203,408){\line(1,-1){60}}
\put(1219,390){\line(1,-1){60}}
\put(1235,372){\line(1,-1){60}}
\put(1251,355){\line(1,-1){60}}
\put(1267,338){\line(1,-1){60}}
\put(1283,322){\line(1,-1){60}}
\put(1299,304){\line(1,-1){60}}
\put(1315,287){\line(1,-1){60}}
\put(1331,270){\line(1,-1){60}}
\put(1347,252){\line(1,-1){60}}
\put(1363,235){\line(1,-1){60}}
\put(1378,219){\line(1,-1){60}}
\put(1394,201){\line(1,-1){60}}
\put(1410,184){\line(1,-1){60}}
%
%
\put(264,158){\line(1,1){60}}
\put(279,174){\line(1,1){60}}
\put(295,190){\line(1,1){60}}
\put(311,209){\line(1,1){60}}
\put(326,225){\line(1,1){60}}
\put(342,241){\line(1,1){60}}
\put(358,258){\line(1,1){60}}
\put(374,275){\line(1,1){60}}
\put(390,293){\line(1,1){60}}
\put(406,311){\line(1,1){60}}
\put(422,327){\line(1,1){60}}
\put(438,344){\line(1,1){60}}
\put(454,362){\line(1,1){60}}
\put(470,379){\line(1,1){60}}
\put(486,396){\line(1,1){60}}
\put(502,413){\line(1,1){60}}
\put(518,430){\line(1,1){60}}
\put(534,448){\line(1,1){60}}
\put(550,465){\line(1,1){60}}
\put(566,482){\line(1,1){60}}
\put(582,499){\line(1,1){60}}
\put(598,517){\line(1,1){60}}
\put(614,534){\line(1,1){60}}
\put(629,550){\line(1,1){60}}
\put(645,567){\line(1,1){60}}
\put(660,583){\line(1,1){60}}
\put(676,600){\line(1,1){60}}
\put(692,618){\line(1,1){60}}
\put(707,634){\line(1,1){60}}
\put(723,650){\line(1,1){60}}
\put(739,668){\line(1,1){60}}
\put(755,685){\line(1,1){60}}
\put(771,702){\line(1,1){60}}
\put(786,719){\line(1,1){60}}
\put(802,735){\line(1,1){60}}
\put(818,753){\line(1,1){60}}
\put(834,770){\line(1,1){60}}
\put(850,787){\line(1,1){60}}
\put(866,804){\line(1,1){60}}
\put(881,820){\line(1,1){60}}
\put(897,838){\line(1,1){60}}
\put(912,854){\line(1,1){60}}
\put(928,871){\line(1,1){60}}
\put(944,888){\line(1,1){60}}
\put(960,905){\line(1,1){60}}
\put(976,923){\line(1,1){60}}
\put(991,939){\line(1,1){60}}
\put(1007,955){\line(1,1){60}}
\put(1023,973){\line(1,1){60}}
\put(1039,991){\line(1,1){60}}
\put(1055,1007){\line(1,1){60}}
\put(1070,1024){\line(1,1){60}}
\put(1086,1040){\line(1,1){60}}
\put(1102,1057){\line(1,1){60}}
\put(1117,1075){\line(1,1){60}}
\put(1133,1091){\line(1,1){60}}
\put(1149,1108){\line(1,1){60}}
\put(1165,1126){\line(1,1){60}}
\put(1181,1142){\line(1,1){60}}
\put(1197,1160){\line(1,1){60}}
\put(1213,1177){\line(1,1){60}}
\put(1229,1194){\line(1,1){60}}
\put(1245,1211){\line(1,1){60}}
\put(1261,1229){\line(1,1){60}}
\put(1277,1246){\line(1,1){60}}
\put(1293,1263){\line(1,1){60}}
\put(1309,1280){\line(1,1){60}}
\put(1325,1298){\line(1,1){60}}
\put(1340,1314){\line(1,1){60}}
\put(1356,1331){\line(1,1){60}}
\put(1372,1348){\line(1,1){60}}
\put(1388,1364){\line(1,1){60}}
\put(1404,1383){\line(1,1){60}}
\put(1420,1400){\line(1,1){60}}
%
%
\put(291,1387){\line(1,0){1117}} \put(319,1356){\line(1,0){1061}}
\put(347,1326){\line(1,0){1005}} \put(375,1297){\line(1,0){950}}
\put(403,1266){\line(1,0){894}} \put(431,1236){\line(1,0){838}}
\put(459,1206){\line(1,0){782}} \put(487,1176){\line(1,0){726}}
\put(515,1146){\line(1,0){670}} \put(542,1117){\line(1,0){615}}
\put(570,1087){\line(1,0){559}} \put(598,1056){\line(1,0){504}}
\put(626,1028){\line(1,0){448}} \put(654,997){\line(1,0){389}}
\put(681,967){\line(1,0){338}} \put(708,940){\line(1,0){283}}
\put(736,909){\line(1,0){228}} \put(764,878){\line(1,0){172}}
\put(791,851){\line(1,0){117}} \put(819,819){\line(1,0){62}}
\put(847,789){\line(1,0){3}} \put(874,762){\line(-1,0){48}}
\put(902,730){\line(-1,0){104}} \put(929,701){\line(-1,0){158}}
\put(957,671){\line(-1,0){214}} \put(985,641){\line(-1,0){270}}
\put(1012,612){\line(-1,0){324}} \put(1040,583){\line(-1,0){380}}
\put(1068,552){\line(-1,0){439}} \put(1095,523){\line(-1,0){493}}
\put(1123,494){\line(-1,0){545}} \put(1151,463){\line(-1,0){601}}
\put(1179,433){\line(-1,0){657}} \put(1207,404){\line(-1,0){713}}
\put(1235,372){\line(-1,0){773}} \put(1263,342){\line(-1,0){829}}
\put(1291,314){\line(-1,0){881}} \put(1319,282){\line(-1,0){941}}
\put(1347,252){\line(-1,0){997}} \put(1374,225){\line(-1,0){1048}}
\put(1402,192){\line(-1,0){1107}}
\end{picture}

\parindent=0.7cm
\parbox{13cm}{
\small Fig.1. Splitting of Minkowski spacetime into submanifolds
stable under Lorentz rotations in $(z,t)$ plain: $R$ and $L$ --
right and left wedges; $F$ and $P$ -- future and past wedges;
$h_{\pm}=h_{\pm}^+\cup h_{\pm}^-$ -- event horizonts; $h_0$ (two
dimential plain $z=t=0$) -- trivial orbit. Variables $\eta$ and
$\rho$ are Rindler coordinates in the right Rindler wedge. The
undushed area is the double Rindler space where Unruh and Fulling
modes coinside and quantization leads to the consept of non --
interacting right and left Fulling particles.}
\end{figure}

Using light cone coordinates $x_{\pm}=t\pm z$ the modes (\ref{boost})
can be represented in the form corresponding to spliting of MS into
domains invariant under Lorentz rotation in $(z,t)$ plain, see Fig.1,
\eq{split}{
\begin{array}{r}\displaystyle
\Psi_{\kappa}(x)=
\theta(x_+)\theta(-x_-)\Psi_{\kappa}^{R}(x)+
\theta(x_+)\theta(x_-)\Psi_{\kappa}^{F}(x)+\\
\displaystyle
+\theta(-x_+)\theta(x_-)\Psi_{\kappa}^{L}(x)+
\theta(-x_+)\theta(-x_-)\Psi_{\kappa}^{P}(x).
\end{array}}
The first term in the r.h.s. of Eq.(\ref{split}) relates to the
right sector of MS, see Fig.1. Nevertheless due to the presence
of Heaviside $\theta$ -- functions it obeys the KG equation with
sources localized on the horizons. Therefore we will consider
the open domain $R$ not containig boundary $h_-^+\cup h_0\cup h_+^+$
-- the Rindler wedge. This manifold is covered by
Rindler coordinates $\eta$, $\rho$
\eq{RC}{z=\rho\cosh\eta,\quad t=\rho\sinh\eta,\quad
-\infty<\eta<\infty,\quad \rho>0,}
and KG equation takes form
\eq{KFG}{
\left\{\frac{\partial^2}{\partial\eta^2}+{\cal K}_R(\rho)\right\}
\phi_R(\xi)=0,\quad
{\cal K}_R(\rho)=-\rho\frac{\partial}{\partial\rho}
\rho\frac{\partial}{\partial\rho}+m^2\rho^2,\quad
\xi=(\eta,\rho).}
Positive frequency solutions of Eq.(\ref{KFG}) with respect to
timelike variable $\eta$ (Fulling modes) are \cite{Fulling}
\eq{Full}{
\Phi_{\mu}(\xi)=(2\mu)^{-1/2} \varphi_{\mu}(\rho) e^{-i\mu\eta},\quad
\varphi_{\mu}(\rho)=
\frac{\sqrt{2\mu\,\sinh\pi\mu}}{\pi}\, K_{i\mu}(m\rho),\quad\mu>0.}
They are orthonormal relative to KG inner product for RS
\eq{KG_R}{\bra f,g\ket_R=i\intR\frac{d\rho}{\rho}\,f^*(\xi)\LRD{\eta}g(\xi).}
Fulling modes $\Phi_{\mu}$ constitute a complete set of positive
frequency solutions for KG equation and therefore may be used for
quantization of the field $\phi_R(x)$ in RS:
\eq{FMquant}{
\phi_R(\xi)=\intR d\mu\,\{\Phi_{\mu}(\xi)c_{\mu}+h.c.\},\quad
[c_{\mu},\hc{c_{\mu'}}]=\delta(\mu-\mu'),\quad
c_{\mu}|0_R\ket=0,\quad \mu>0,}
where the state $|0_R\ket$ is called Fulling vaccum.
Annihilation operators of Fulling particles $c_{\mu}$ may be
expressed in terms of the field $\phi_R(\xi)$ by
\eq{c}{
c_{\mu}=\bra\Phi_{\mu},\phi_R\ket_R=
\frac{i}{\sqrt{2\mu}}
\intR\frac{d\rho}{\rho}\varphi_{\mu}(\rho)
\left.\left[\Diff{\eta}\phi_R(\xi)-
i\mu\phi_R(\xi)\right]\right\vert_{\eta=0}}
Crucial point for the concept of Fulling particles is the requirment for
the field $\phi_R(\xi)$ to obey the boundary condition \cite{last} at
$\rho=0$
\eq{bc}{\lim_{\rho\to 0}\phi_R(\rho,\eta)=0,}
besides a trivial null condition at $\rho=\infty$. Because of unboundedness
of the operator $\phi_R(\eta,\rho)$ the relation (\ref{bc}) should be
understood as a condition for matrix elements between physically
realizable states
\footnote{By physically realizable states we mean the states
corresponding to finite mean values of operators
$H_R=\intR d\mu\,\mu \hc{c_{\mu}}c_{\mu}$ and $H_R^{-1}$. The second
requirment arises due to absence of the mass gap for Fulling particles,
see, e.g., sec. 4 of Ref.\cite{Kay}.}. It is worth noting that the
substitution $\rho=m^{-1}e^u$ mapping the point $\rho=0$ into $\rho=\infty$
reduces the condition (\ref{bc}) to the usual requirment for vanishing
of the field at spatial infinity.

\parindent 0.7cm
\newsec
To perform in MS quantization similar to Fulling one in RS Unruh
suggested \cite{Unruh,UW} to use modes which are superpositions
of boost modes with positive and negative frequencies. In notation
of Ref.\cite{last} they read
\eq{Unruh_Modes}{
R_{\mu}=\frac
{\left[ e^{\pi\mu/2}\Psi_{\mu}-e^{-\pi\mu/2}\Psi_{-\mu}^*\right]}
{\sqrt{2\sinh\pi\mu}},\quad
L_{\mu}=\frac
{\left[e^{\pi\mu/2}\Psi_{-\mu}^*-e^{-\pi\mu/2}\Psi_{\mu}\right]}
{\sqrt{2\sinh\pi\mu}},\quad \mu>0.}
Although Unruh modes $R_{\mu}$ ($L_{\mu}$) are not positive frequency with
respect to Minkowski time $t$ they are positive (negative) frequency
in $R$ and $L$ wedges of MS with respect to timelike variables
$\half\ln(\pm x_+/\mp x_-)$ respectively. Unruh modes have remarkable
properties
\eq{props}{
\begin{array}{l}
\displaystyle
R_{\mu}(x)= 0\;\mbox{in}\;L,\quad
R_{\mu}(x)= \Phi_{\mu}(x)\;\mbox{in}\;R,\quad
\bra R_{\mu}, R_{\mu'}\ket_{\M}=\delta(\mu-\mu'), \\
\displaystyle
L_{\mu}(x)\equiv 0\;\mbox{in}\;R,\quad
L_{\mu}(x)\equiv \Phi_{\mu}(-x)\;\mbox{in}\;L,\quad
\bra L_{\mu}, L_{\mu'}\ket_{\M}=-\delta(\mu-\mu').
\end{array}}
Note that $R$ -- modes are not analitical continuation of Fulling
modes (\ref{Full}) because different rules are used for continuation
of $\Psi_{\mu}(x)$ and $\Psi_{-\mu}^*(x)$ when passing around branch
points $x_{\pm}=0$. Inverting Eqs.(\ref{Unruh_Modes}) and substituting
the result into Eq.(\ref{BMquant}) one obtains for $x\ne 0$ (when it
is possible to split the integral in Eq.(\ref{BMquant}) into two
integrals over $\kappa>0$ and $\kappa<0$)
\eq{Uquant}{
\phi(x)=\intR d\mu\,\{R_{\mu}(x) r_{\mu}+ L_{\mu}(x) \hc{l_{\mu}}+h.c.\},}
where operators
\eq{rl}{
\begin{array}{c}\displaystyle
r_{\mu}=\frac
{\left[ e^{\pi\mu/2}b_{\mu}+e^{-\pi\mu/2}\hc{b_{-\mu}}\right]}
{\sqrt{2\sinh\pi\mu}},\quad
l_{\mu}=\frac
{\left[e^{\pi\mu/2}b_{-\mu}+e^{-\pi\mu/2}\hc{b_{\mu}}\right]}
{\sqrt{2\sinh\pi\mu}},\quad \mu>0, \\
\displaystyle
[r_{\mu},\hc{r_{\mu'}}]=
[l_{\mu},\hc{l_{\mu'}}]=
\delta(\mu-\mu').
\end{array}}

Unfortunately Eqs.(\ref{Uquant}),(\ref{rl}) does not define valid
quantization in the whole MS because splitting of the integral over
$\mu$ into separate terms describing creation and anihilation parts
of the field cannot be performed in $F$ and $P$ wedges. Indeed,
using explicit expression for boost modes \cite{last} one finds
\eq{sing}{
R_{\mu}^F r_{\mu}=-L_{\mu}^F \hc{l_{\mu}} =
-\frac{i}{2\sqrt{2\pi}\mu} J_0(m\sqrt{x_+x_-})\,\phi(0,0),\quad \mu\to 0,}
where according to Eqs.(\ref{boost}), (\ref{BMquant})
\eq{FIzero}{\phi(0,0)=\frac1{\sqrt{2}}(b_0+\hc{b_0})\ne 0.}
Hence the integral over $\mu$ of these expressions diverges
logarithmically on the low limit while in the sum of these terms
in Eq.(\ref{Uquant}) singularities cancel.
Of course the Eqs.(\ref{sing}), (\ref{FIzero}) should be understood
as relation between matrix elements, see discussion after Eq.(\ref{bc}).

Using Eqs.(\ref{BMquant}), (\ref{rl}) we obtain the relation
\eq{main2}{
\bra 0_{\M}|\hc{r_{\mu}}r_{\mu'}|0_{\M}\ket=
\frac1{\exp(2\pi\mu)-1}\delta(\mu-\mu').}
The Eq.(\ref{main2}) is usually interpreted as a special case of
Eq.(\ref{main1}) for $\Robs=\hc{r_{\mu}}r_{\mu'}$ under assumption that
$r_{\mu}=c_{\mu}$. Moreover Eq.(\ref{main2}) is sometimes used for
derivation of the general result (\ref{main1}) and thus is equivalent
to it. Nevertheless such interpretation is not valid since according to
Eqs.(\ref{Uquant}),(\ref{sing}) the operators $r_{\mu}$, $l_{\mu}$,
$\hc{r_{\mu}}$, $\hc{l_{\mu}}$ can not be considered as anihilation and
creation operators of Fulling - Unruh particles in MS where the vaccum state
$|0_{\M} \ket$ is defined. This is not a surprise since it is impossible
to find any time -- like variable in MS relative to which the Unruh modes
(\ref{Unruh_Modes}) correspond to frequency of definite sign.
Quantization of the field in the basis of Unruh modes could be performed
only if the boundary condition
\eq{BC}{\lim_{z\to 0} \phi(0,z)=0}
existed. The condition (\ref{BC}) is equivalent to the boundary condition
for the field in RS. Hence quantization of the field in the basis of Unruh
modes can be performed only in double RS (a disjoint union of wedges $R$
and $L$, see Fig.1) rather than in MS and the r.h.s. of the
Eqs.(\ref{main1}),(\ref{main2}) can not be considered as thermal equilibrium
expectation values.

\parindent 0.7cm
\newsec
Let us turn back now to the discussion of Eq.(\ref{main1}) which
encounter mathematical difficulties in the conventional formalism of
quantum field theory. The representation of cannonical commutation
relations in terms of Unruh operators (\ref{rl}) is unitary inequivalent
to the one in terms of operators $b_{\kappa}$, see Eq.(\ref{BMquant}).
It is a direct consequence of divergency of $Z$ in Eq.(\ref{main1})
\cite{Fulling}. There are two ways to avoid this difficulty. The first
one is to place the field in the box which may in this problem be
constructed by two uniformly accelerated mirrows moving in right
and left Rindler wedges \cite{CR}. However such regularization again
leads to consideration of double RS as a physical spacetime of
the observer. The second opportunity is to use algebraic approach and a
notion of KMS state as a definition of thermal equilibrium state.

To reformulate the Eq.(\ref{main1}) in terms of algebraic approach
let us introduce the required definitions.
Let $D$ be a linear symplectic space of solutions of KG equation
for $C^{\infty}$ Cauchy data with compact support on some
surface $\KS$ (a classical phase space of field theory). An algebra
$\Alg$ of observables of the field is a $C^*$ algebra with generators
$W(\Phi),\; \Phi\in D$ satisfying usual Weyl relations
\eq{CCR}{
W(\Phi_1)W(\Phi_2)=\exp(-i\spr{\Phi_1}{\Phi_2}/2)W(\Phi_1+\Phi_2),\quad
W(\Phi)^*=W(-\Phi)}
with $\sigma$ being a symplectic product on $D$. The states in algebraic
approach are linear functionals on $\Alg$. Vacuum state and the
corresponding representation of $\Alg$ by operators acting in Hilbert
space can be constructed by using a one particle structure which maps
$D$ onto the Hilbert space of physically realizable positive frequency
solutions
\footnote{In this construction the space $D$ plays a role of a
complete set of quantum numbers in the usual formalism.}.
Let $D_R$, $D_L$ be subspaces of $D$ consisting of those solutions which
vanish in closed wedges $\bar L$, $\bar R$ respectively and
$\tilde D=D_L\oplus D_R\subset D$. Note that $\tilde D$ is a subspace of
those solutions which vanish in a neighourhood of $h_0$, see Fig.1.
Finite linear combinations of elements from $\Alg$ of the form $W(\Phi)$
with $\Phi\in D_R$ constitute an open (rather than $C^*$) subalgebra
$\RWA$ of $\Alg$ which is called the right wedge algebra. The left
wedge algebra $\LWA$ and the double wedge algebra $\DWA$ are defined
similarly by changing $D_R$ to $D_L$ and $\tilde D$ respectively.

According to the Bisognano-- Wichmann theorem \cite{BW} the Minkowski
vacuum state $\omega_{\M}$ when restricted to the right wedge
algebra $\RWA$ satisfies KMS condition with temperature
$(2\pi)^{-1}$ with respect to boost time $\eta$.
But in order to give a physical interpretation to this theorem
one must relate it to the procedure of measurement by pointing out
the quanta which are thermally distributed. Such interpretation
is given by the notion of double KMS state.

Let $\W_{\M}$ be usual Minkowski vacuum state and $\tilde\W_F^{(2\pi)}$
be a $(2\pi)^{-1}$-- temperature double KMS state over "double linear
system" $\DWA=\LWA\otimes\RWA$ with respect to the boost evolution
on $\RWA$ and "antiboost" evolution on $\LWA$ (see Ref.\cite{Kay},
section 1 for exact definitions and eqs.(2.9)--(2.11) for
explicit construction). The main result (proposition 2.1 of Ref.\cite{Kay})
is that
\eq{Unruh}{\W_{\M}=\tilde\W_F^{(2\pi)}\quad \mbox{on $\DWA$}.}
This equation is an analog of eq.(\ref{main1}) in algebraic approach.

Let us explicitly evaluate eq.(\ref{Unruh}) and check that it can
not be extended to the whole algebra of observables of the free
field $\Alg$. In the case of theory in MS an expectation value of
Weyl generator in Minkowski vacuum state is defined by \cite{Kay}
\eq{MV}{\W_{\M}(W(\Phi))=\exp\left(-\half ||K_{\M}\Phi||^2\right),}
where  $K_{\M}$ is a ground one particle structure for this case
which is a map extracting a positive frequency part of the
solution, $K_{\M}:\Phi\mapsto\Phi^{(+)}$ and \eq{normaM}{
||K_{\M}\Phi||^2=\bra\Phi^{(+)},\Phi^{(+)}\ket_{\M}= \intRR
d\kappa\, |\bra\Psi_{\kappa},\Phi\ket_{\M}|^2,} where we have used
a complete set of boost modes $\Psi_{\kappa}$ (\ref{boost}) to
extract positive frequency part. By inverting relations
(\ref{Unruh_Modes}) one can rewrite eq.(\ref{normaM}) in terms of
Unruh modes. The result is
\eq{normaM1}{
 ||K_{\M}\Phi||^2= \intR
d\mu\, \coth\pi\mu  \, \{ |\bra R_{\mu},\Phi\ket_{\M}|^2+|\bra
L_{\mu},\Phi\ket_{\M}|^2\} +...} (here and below dots denote the
correlation term).

Now let us evaluate expectation value of Weyl generator in a double
KMS state with temperature $\beta^{-1}$ with respect to Fulling
ground one particle structure. This expectation value may be
written as \cite{Kay} \eq{MV1}{\tilde\W_F^{(\beta)}(W(\Phi))=
\exp\left(-\half ||\tilde K_{F}^{(\beta)}\Phi||^2\right),} where
$\tilde K_F^{(\beta)}$ is a double KMS one particle structure.
According to definition of $\tilde K_F^{(\beta)}$ as a map into a
direct sum of two copies of Hilbert space of positive frequency
solutions in RS one has
\eq{normaKMS}{||\tilde
K_{F}^{(\beta)}\Phi||^2= ||K_F\Phi_R||_{\beta}^2+ ||K_F\,{\cal F}
\Phi_L||_{\beta}^2 +...,}
where $\Phi_{R,\, L}$ is the restriction
of $\Phi$ to the right (left) Rindler wedge, ${\cal
F}\Phi_L(\xi)=\Phi_L(-\xi)$. Quantity $||K_F\Phi_R||_{\beta}^2$ is
defined by \footnote{We refer to sect. 1.4 of Ref.\cite{Kay} for
detailed explanations on construction of the map $\tilde
K_F^{\beta}$.}: \eq{therm_norm}{ ||K_F\Phi_R||_{\beta}^2=
\int\limits_{\KS\cap R} d\sigma_{\xi} \int\limits_{\KS\cap R}
d\sigma_{\xi'} \,\WF_F^{(\beta)}(\xi,\xi')
\frac1{\rho}\LRD{\eta}\Phi_R(\xi)\frac1{\rho'}\LRD{\eta'}\Phi_R(\xi'),}
where $\WF_F^{(\beta)}(\xi,\xi')$ is the thermal Whightman function
for theory in RS and Cauchy surface $\KS$ is choosen to be
$\eta=0$. For explicit calculation let us use a complete set of
Fulling modes (\ref{Full}). One can express thermal Whightman
function in terms of Fulling modes by \eq{ThermWF}{
\WF^{(\beta)}(\xi,\xi')=\intR \frac{d\mu}{\exp(\beta\mu)-1}\,
\{\Phi_{\mu}(\xi)\Phi_{\mu}^*(\xi')\exp(\beta\mu)+
\Phi_{\mu}^*(\xi)\Phi_{\mu}(\xi')\}.} Using Eqs.(\ref{therm_norm})
-- (\ref{ThermWF}) one obtaines \eq{norma_res}{ ||\tilde
K_F^{(\beta)}\Phi||^2= \intR d\mu\,\coth\left( \frac{\beta\mu}2
\right)\, \{ |\bra\Phi_{\mu},\Phi_R\ket_R|^2+ |\bra\Phi_{\mu},{\cal
F}\Phi_L\ket_R|^2 \} +...}

Taking into account the relation \cite{last} \footnote{The second
term is responsible for canceling of divergent part of
$\bra\Phi_{\mu},\Phi_R\ket_R$ if $\Phi(0,0)\ne 0$. Note that we
improved in Eq.(\ref{rel}) an obvious misprint made in Eq.(20) of
the Ref.\cite{last}.} \eq{rel}{
\begin{array}{r} \displaystyle
\bra R_{\mu},\Phi\ket_{\M}=\bra\Phi_{\mu},\Phi_R\ket_R+
\frac{i}{2\pi}\sqrt{\sinh\pi\mu}\,\lim_{z\to 0}\Phi(0,z) \times \\
\displaystyle
\times
\left\{\Gamma(i\mu)\left(\frac{mz}2\right)^{-i\mu}-
\Gamma(-i\mu)\left(\frac{mz}2\right)^{i\mu}\right\}
\end{array}}
one concludes after comparing Eqs.(\ref{norma_res}), (\ref{normaM1})
that equation
\eq{main}{\W_{\M}(W(\Phi))=\tilde\W_F^{(2\pi)}(W(\Phi))}
(and hence by linearity Eq.(\ref{Unruh})) holds if and only if $\Phi(0,0)=0$
or in other words only for $\Phi\in\tilde D$.

We see that Eq.(\ref{Unruh}) holds only on the dense subalgebra
$\DWA\subset \Alg$, which corresponds to the space of those solutions
for the field equation which satisfy boundary condition at the plain $h_0$.
Therefore the l.h.s. of Eq.(\ref{Unruh}) admits continuation to the whole
$\Alg$ while the r.h.s. does not.

Let us consider two opportunities to interpret Eq.(\ref{Unruh}).
The first one is to treat $\Alg$ as the true algebra of observables
for the accelerated observer. In this case Eq.(\ref{Unruh}) does not
hold for all observables and therefore Minkowski vacuum does not coincide
with the thermal state $\tilde\W_F^{(2\pi)}$.

The second opportunity is to insist that $\DWA$ should be the true algebra
of observables for accelerated observer (although this is not $C^*$
algebra as it is usually required). In this case Poincar\'e invariance
is lacking and hence the true Minkowski vacuum state $\W_{\M}$ does not
exist. Then Eq.(\ref{Unruh}) is satisfied for all physical observables
and hence the {\it restriction} $\W_{\M}\vert_{\DWA}$ of the state
$\W_{\M}$ to $\DWA$ coincides with the state $\tilde\W_F^{(2\pi)}$ and
admits interpretation in terms of Fulling -- Unruh quanta. But the
restriction $\W_{\M}\vert_{\DWA}$ of Minkowski vacuum state to the
subalgebra $\DWA$ is not Minkowski vacuum state any more because of
the other domain of definition.


\parindent 0.7cm
\newsec
Presence of boundary condition (\ref{BC}) shows that Unruh quantization
can not be performed in MS. From the physical point of view the meaning
of Eq.(\ref{BC}) is that the plain $h_0$ does not affect any physically
realizable measurements and therefore should be considered as being
removed from the spacetime. But such removal crucially changes topological
properties
\footnote{The point of view that Unruh effect arises if the true spacetime
is $\M\backslash h_0$ has been previously discussed in
Refs.\cite{CrDuff},\cite{Troost}.}
and symmetry group
\footnote{Lacking of Poincar\'e invariance for the field theory with boost
time evolution is a consequence of the fact that boost transformations
do not constitute a normal subgroup in Poincar\'e group.}
of the spacetime. Therefore the known general relation (\ref{Unruh}) between
Minkowski vacuum state and thermal Fulling -- Unruh state is by no way
physically related to the behaviour of accelerated detector in empty MS.

A separate aspect of Unruh problem is weather a concrete accelerated
detector with known structure would behave as if having been
immersed in thermal bath with Davies -- Unruh temperature. A rather
complete treatment of this problem was done \cite{NR} for the case when
elementary particles are used as detectors and constant electric field is
employed as accelerating force. It occured that only for some set of values
of parameters of the model such detectors demonstrate Unruh type
behaviour.
An example of utilization of a composite system (a heavy atom or ion)
as an accelerated detector was considered nonrelativistically in
Ref.\cite{MKP}. It was shown that due to the tunneling ionization process
in accelerating electric field such detector will be destroyed long
before it comes to thermal equilibrium state with Davies -- Unruh
temperature.

Because a systematic relativistic theory of bound states has not been
created yet a question of behaviour of accelerated detector in general
case is still open.

\vspace{1cm}
\centerline{\bf Acknowledgements}

N.B. Narozhny and A.M. Fedotov are grateful to Dr. R. Ruffini for
hospitality at Rome University "La Sapienza" (Italy). V.A. Belinskii
thanks the Institut des Hautes Etudes Scientifiques at Bures -- sur --
Yvette (France) where a part of the work for this paper was done
for hospitality and support.
This work was supported in part by the Russian Fund for Fundamental
Research under projects 97--02--16973 and 98--02--17007.

\vglue 0.5cm
\centerline{\bf\large References.}\nopagebreak
{
}

\end{document}